# Flow Simulation of Lid-Driven Rectangular Cavity by Using Lattice Boltzmann Method

Xiuqiao Xiang [a, c], Baochang Shi [b*]

[a] School of Computer Science, China University of Geosciences, Wuhan 430078, PR China
[b] School of Mathematics and Statistics, Huazhong University of Science and Technology, Wuhan 430074, PR China
[c] National Engineering Research Center of Geographic Information System, China University of Geosciences, Wuhan 430078, PR China

**Abstract**: Wall-driven flow in square cavity has been studied extensively, yet it is more frequently for the rectangular cavity flow occurring in practical problems, and some flow characteristics about rectangular cavity have not been fully investigated. As a promising numerical simulation tool, the Lattice Boltzmann Method (LBM) is employed to simulate the lid-driven flow in a two-dimensional rectangular cavity in this paper. First, the code is validated for the standard square cavity, the velocity profiles, stream function values and center positions of the primary and second vortexes at different Re are presented and compared with previous researches. Then, the eddy dynamics of rectangular cavities is simulated and discussed with Reynolds number (Re) in the range of 4000-8000 and vertical to horizontal axis ratio (Ar) varied from 0.4 to 2.0, and the streamline and center migration of the primary vortex are drawn realistically for rectangular cavity. In the end, the steady, periodic, aperiodic and unstable phenomenon in the rectangular cavity is produced by LBM, Re and Ar which play the significant role in the state transition of lid-driven rectangular cavity flow are analyzed and summarized in detail. By the LBM numerical simulation with $500 \times (500*Ar)$ grid, we have discovered that the flow state in the rectangular cavity with a fixed Ar changes monotonically from stable state, periodic state to aperiodic and unstable state with the increase of Re, while the flow state in the rectangular cavity with a fixed Re varies non-monotonically with the increase of Ar. Additionally, the cycle length of periodic states varies with the change of Ar and Re.

## 1. Introduction

Lid-driven flow in a two-dimensional square cavity is a classical, benchmark problem where the phenomena of closed flow could be seen in industrial field, such as chemical etching or film coating [1, 2]. The classic flow problem has been investigated since 60's decade and developed very fast with the rapid development of computer science. In the last decades, a variety of powerful numerical methods, including finite difference, finite element, finite volume, spectral method and variation iteration method [3-7], have been intensively studied, and the simulation results are shown to agree quite well with the corresponding solutions of Navier–Stokes (NS) equations. Some standardized conclusions about the square cavity flow have been obtained, which can be used to validate new numerical methods [8]. Beside conventional numerical methods listed above, the lattice Boltzmann Method (LBM) is being successfully used to the numerical simulation of fluid flow. The LBM has many advantages, such as the simplicity of program, locality of computation, natural parallelism and easiness in dealing with complex boundary [9, 10]. An introduction to LBM theory, its methodology and the current status may be obtained from Refs. [11-13].

By performing a detailed analysis of the lid-driven square cavity flow, the LBM simulation results are compared excellently with the conventional (NS) results, which strongly confirms the capabilities of LBM [14, 15]. Lemée et al. (2015) reported that the symmetric flows is stable and steady for Reynolds number under 4000 (the Reynolds number is defined as $\text{Re} = UL/\nu$, where U is the lid-velocity, L the cavity-width, and $\nu$ the kinematic viscosity of the enclosed fluid in motion) [16]. Despite the fact that the geometry of the lid-driven square cavity is very simple, the fluid flow retains a rich physics character manifested by multiple counter-rotating recirculating regions on the corners of the cavity, several features of the flow, such as the streamlines, the location and strength of the primary vortex, and the eddy dynamics are addressed and compared with previous findings from experiments and theory [17-19]. Since the realistic cavity flows are not limited in the rectangular cavity flow, lid-driven flows in complex geometries, such as the flows in a skewed, semi ellipse, L-Shaped, triangular or trapezoidal cavity, are discussed, the vortex structure and its evolutionary features varied with Re and D are analyzed in detail [20–25]. Wall-driven cavity with two opposing walls moving at the same speed and the same or opposite directions has been studied in the literature [26], extensive numerical simulations have been undertaken to gain approximate ranges for the critical Hopf and Neimark-Sacker bifurcations for the classic and two two-sided cavity configurations. The threshold for transition to chaotic motion is also reported. To observe the effect of various Reynolds number and size of the internal circular obstacles on the flow characteristics and primary, secondary vortex formation, the LBM has been applied for the simulation of lid-driven flows inside a square cavity with internal circular obstacles of various diameters under Reynolds numbers ranging from 100 to 5000 [27]. These works available attain important achievements on the lid-driven flow of square cavity.

*Corresponding author, e-mail address: shibc@hust.edu.cn, xiangxq0005@126.com
Tel: 86-027-87543231, 86-027-67883716, Fax: 86-027-87543231

Currently, the simulation of lid-driven flow is studied extensively, and most work is focused on the flow in a square cavity with aspect ratio Ar = (H/L)=1, H being the cavity-depth and L the width. However, in reality it's rarely the case where aspect ratio Ar is exactly equal to 1, it was found that, the flow in a rectangular cavity with aspect ratio Ar≠1 (Ar is less or greater than 1) has a wider field of applications compared with the square cavity flow. Therefore, from then on many scholars began to study the rectangular cavity flow. In Ref. [28] steady results are presented for deep cavities with aspect ratios of 1.5-4, and Reynolds numbers of 50-3200. Several features of the flow, such as the location and strength of the primary vortex, and the corner-eddy dynamics are explored and compared with previous findings. Steady results for deep cavities show the existence of corner eddies at the bottom, which coalesce to form a second primary-eddy as the cavity aspect-ratio is increased above a critical value. However, at relatively high Reynolds numbers, the second primary-eddy is formed via a rapid transition of an unsteady wall-eddy. From Ref. [29] the detailed analysis of the vortex structure is carried out and its evolutionary features for the rectangular cavity flow are provided, which demonstrates that there are significant differences for the rectangular cavity flow with various aspect ratios (Ar) and Reynolds number (Re). In Ref. [30] the characterization of oscillatory instability is analyzed in a lid-driven rectangular cavity. The numerical results achieved above attract worldwide attention. However, there existing some work on driven flows in rectangular cavity worthy of in-depth study. Therefore, the aim of this paper is to extend the LBM simulations for rectangular 2D cavities with aspect ratio $0.4 \leq Ar \leq 2.0$ and $4000 \leq Re \leq 8000$, specifically to discover the interesting phenomena and analyze the state transition of rectangular cavity varied with Re and Ar by investigating the streamline, center location of the vortexes, the flow speed $U_x$, $U_y$, and $U_x$-$U_y$ phase diagram of observing points.

The rest of this paper is organized as follows. In Section 2, the mathematical formulation of LBM and its processing strategy including the initial condition, boundary condition are described. In Section 3, LBM simulation is performed on the two-dimensional lid-driven rectangular cavity with different Ar and Re, and the numerical results are presented and analyzed. Finally, the major conclusions are summarized in Section 4.

## 2. LBM and its processing strategy

Before the LBM numerical simulation of fluid flows, the lattice Boltzmann mathematical model together with its experimental processing strategy including boundary conditions, initial conditions are described in this section.

### 2.1 LBM Mathematical formulation

LBM based on DnQb lattice with b velocity directions in n-dimensional space can be effectively applied to simulate evolution equations [11, 12]. If we define $f_j(x,t)$ as the particle distribution function of the species with velocity $c_j$ at time $t$ and position $x$, the evolution equations of the particle distribution function in LBM are given by

$$f_j(x+c_j\Delta t, t+\Delta t) = f_j(x,t) - \frac{1}{\tau}\left[f_j(x,t) - f_j^{eq}(x,t)\right] \tag{1}$$

where $\Delta t$ is the time step, $\tau$ represents the dimensionless relaxation time, $f_j^{(eq)}(x,t)$ is the equilibrium distribution function

$$f_j^{eq} = \omega_j \rho \left[1 + \frac{c_j \cdot u}{c_s^2} + \frac{(c_j \cdot u)^2}{2c_s^4} - \frac{u \cdot u}{2c_s^2}\right] \tag{2}$$

where $\rho$ is the fluid density, u=($u_x$, $u_y$) is the fluid speed, the sound speed $c_s$ is related to the particle speed $c$, $c_s^2 = c^2/3$. $c = \Delta x/\Delta t$ and $\Delta x$ is the lattice spacing. $\{c_j, j=0,..., b-1\}$ is the set of discrete velocity directions. For the D2Q9 model commonly used in LBM, their corresponding discrete velocity set $\{c_j, j=0,\cdots,8\} = \{(0,0),(\pm c,0),(0,\pm c),(\pm c,\pm c)\}$, and weight coefficients $\omega_0 = 4/9$, $w_{1-4} = 1/9$, $\omega_{5-8} = 1/36$.

The local and instantaneous macroscopic flow properties ($\rho$, u) are related to the local and instantaneous distribution function in the following manner:

$$\begin{aligned}\rho(x,t) &= \sum_{j=0}^{8} f_j(x,t) \\ \rho u(x,t) &= \sum_{j=0}^{8} c_j f_j(x,t)\end{aligned} \tag{3}$$

To recover the macroscopic Navier–Stokes equations (see Eq.(4)) correctly in the limit of low Mach number, Taylor expansion, Chapman–Enskog expansion and multi-scale technique are utilized to Eq.(2) in succession.

$$\begin{aligned}\partial_t \rho + \nabla \cdot (\rho u) &= 0 \\ \partial_t (\rho u) + \nabla \cdot (\rho u u) &= -\nabla p + \nabla \cdot \left[\rho v(\nabla u + (\nabla u)^T)\right]\end{aligned} \tag{4}$$

where $p$ is the pressure, and $v$ is the kinematic viscosity, which are, respectively, given by

$$\begin{aligned}p &= \rho c_s^2 \\ v &= c_s^2 (\tau - \frac{1}{2})\Delta t\end{aligned} \tag{5}$$

## 2.2 LBM experimental processing strategy

(i) Initial conditions

Initially, at the time of starting the simulations, $t=0$, the velocities ($u_x$ and $u_y$) at all the nodes are set to be zero 0 except the nodes at top wall $u_x$=0.1, the initial density $\rho$ is taken as 1.0. The initial distribution function $f_j(x,0)$ is equal to the initial equilibrium distribution function $f_j^{(eq)}(x,0)$ that is evaluated according to Eq.(2) and the initial conditions listed above.

(ii) Boundary conditions

To simulate the steady flow in a rectangular cavity, it is necessary to choose an appropriate approach to deal with the boundary conditions in the LBM. The non-equilibrium extrapolation scheme for boundary treatment [31] has been widely applied in the simulations due to its second-order accuracy and good numerical stability. Therefore, this boundary condition is applied to the boundaries of the rectangular cavity in the following simulations. In addition, for the rectangular cavity flow problem, boundary conditions are used and described as follows:

Top wall: $u_x$=0.1, $u_y$=0;
Other walls: $u_x$=0, $u_y$=0.

(iii) Convergence criterion

The convergence criterion of LBM experiment is given as follows:

$$\sqrt{\frac{\sum_{i,j}\left\{\left[u_x(i,j,t+100\Delta t)-u_x(i,j,t)\right]^2+\left[u_y(i,j,t+100\Delta t)-u_y(i,j,t)\right]^2\right\}}{\sum_{i,j}\left[u_x^2(i,j,t)+u_y^2(i,j,t)\right]}}<10^{-6} \quad (6)$$

The simulation is considered to have reached a steady state when the velocity change in ($u_x$, $u_y$) values at two time-levels separated by 100 time steps is less than $10^{-6}$.

## 3. Numerical experiments

In this section, we study the lid-driven flow in a rectangular cavity by using LBM mentioned above and perform numerical experiments related. The particle speed $c$=1.0, $\Delta x = \Delta t$. The top wall is shifted from left to right with a constant velocity $u$=0.1.

### 3.1 Code Validation of the LBM

In order to validate the in-house code, in this subsection, we use the LBM to simulate the lid-driven square cavity flow (Ar=1) in a resolution of 256×256, and compare the simulation results obtained by the LBM with those published data [3] [9]. The velocity profiles along the vertical central line, at $x$=0.5, and the horizontal line, at y= 0.5, are exhibited in Fig.1, in which the numerical results calculated by the present LBM agree well with the data produced by the operator-splitting, finite elements method [3].

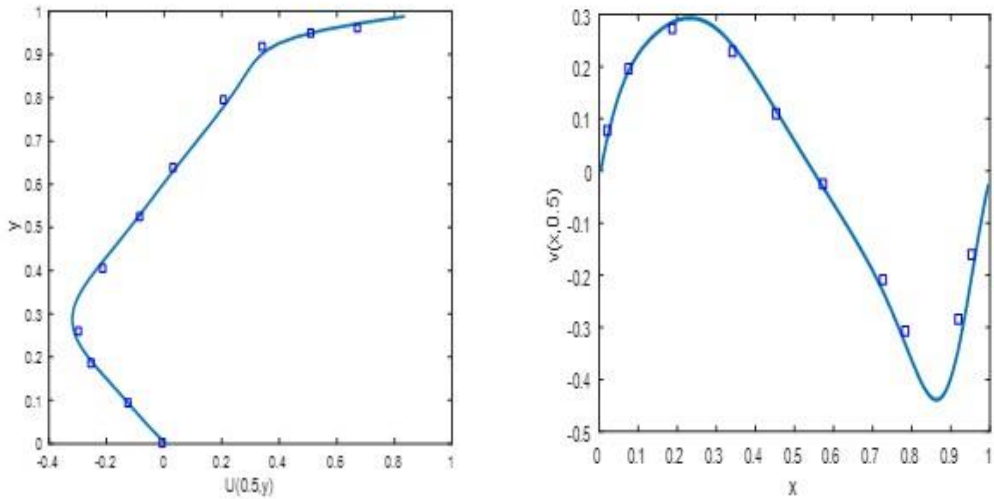

a) The vertical central line, at $x$=0.5       b) The horizontal central line, at $y$=0.5

Fig.1 The velocity profiles along the central line compared with the previous work

Solid lines correspond to the results by the present LBM, □ denote the points from Ref.[3]

Then, we use LBM to simulate the lid-driven square cavity flow (Ar=1) in a resolution 500×500, the velocity profiles along the central line are presented in Fig.2, which validates that the simulation results by the present LBM with resolution 500×500 coincide well with the data produced by LBM in a resolution of 256×256, this reinforces the applicability of the LBM used for the following numerical experiments in a resolution 500×(500*Ar).

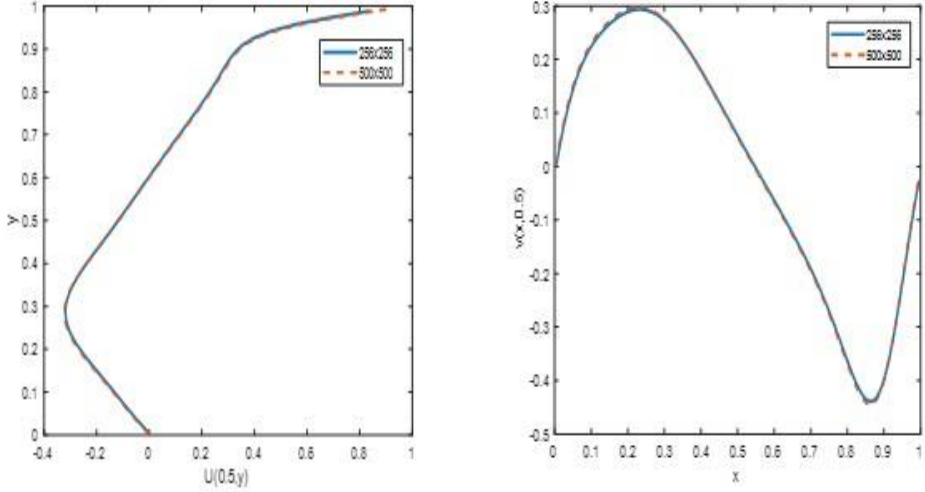

a) The vertical central line, at $x$=0.5  b) The horizontal central line, at $y$=0.5

Fig.2 The velocity profiles along the central line at the different grid 256×256 or 500×500

In addition, in order to further check the performance of LBM used, the stream-function ($\Psi$) value and the location of primary vortex and secondary vortex at different Reynolds number (Re) are summarized in Table 1, from which the simulation results obtained by LBM also agree well with those results of Refs. [3] [9]. Therefore, the LBM has the capability to simulate the rectangular cavity flow.

Table 1 Stream function values and center positions of the primary and secondary vortexes for different Re

| Re | | Primary vortex | | | lower left vortex | | | lower right vortex | | |
|---|---|---|---|---|---|---|---|---|---|---|
| | | $\Psi_{max}$ | x | y | $\Psi_{min}$ | x | y | $\Psi_{min}$ | x | y |
| 400 | a | 0.1139 | 0.5547 | 0.6055 | -1.42e-5 | 0.0508 | 0.0469 | -6.42e-4 | 0.8906 | 0.1250 |
| | b | 0.1121 | 0.5608 | 0.6078 | -1.30e-5 | 0.0549 | 0.0510 | -6.19e-4 | 0.8902 | 0.1255 |
| | c | 0.1113 | 0.5547 | 0.6055 | -1.2501e-05 | 0.0508 | 0.0469 | -5.9418e-04 | 0.8867 | 0.1211 |
| 1000 | a | 0.1179 | 0.5313 | 0.5625 | -2.31e-4 | 0.0859 | 0.0781 | -1.75e-3 | 0.8594 | 0.1094 |
| | b | 0.1178 | 0.5333 | 0.5647 | -2.22e-4 | 0.0902 | 0.0784 | -1.69e-3 | 0.8667 | 0.1137 |
| | c | 0.1142 | 0.5313 | 0.5664 | -1.8389e-4 | 0.0820 | 0.0742 | -1.60e-3 | 0.8633 | 0.1094 |
| 5000 | a | 0.1190 | 0.5117 | 0.5352 | -1.36e-3 | 0.0703 | 0.1367 | -3.08e-3 | 0.8086 | 0.0742 |
| | b | 0.1214 | 0.5176 | 0.5373 | -1.35e-3 | 0.0784 | 0.1373 | -3.03e-3 | 0.8078 | 0.0745 |
| | c | 0.1030 | 0.5156 | 0.5586 | -1.80e-04 | 0.0781 | 0.1406 | -2.9e-3 | 0.8086 | 0.0742 |

Note: a, Ghia U [3]; b, Hou S [9]; c, present work.

### 3.2 Streamline and vortex structure of the rectangular cavity flow at different cases

Here and later in this paper, we use the LBM to simulate the rectangular cavity flow with different Ar (0.4≤Ar≤2.0) and Re (4000≤Re≤8000) under GPU computing environment, most of the experimental simulations were carried out on 500×(500*Ar) lattice nodes by performing code execution four million steps at least.

### (i) The rectangular cavity flow with fixed Ar, Re while different evolutional steps

Fig.3 presents the streamlines for the rectangular cavity flow with fixed Ar=1.2, Re=6000 at different instants of time (evolution step), which shows that a counter-rotating vortex is formed below the moving lid, a wall-eddy is produced in the rectangular cavity with the increasing of evolution time (step). Furthermore, more and more eddies are observed at the bottom of the cavity with the extension of the evolution time.

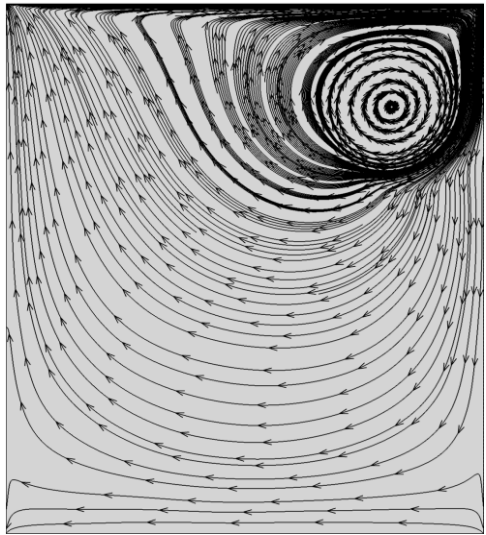
(a) 20000 Step

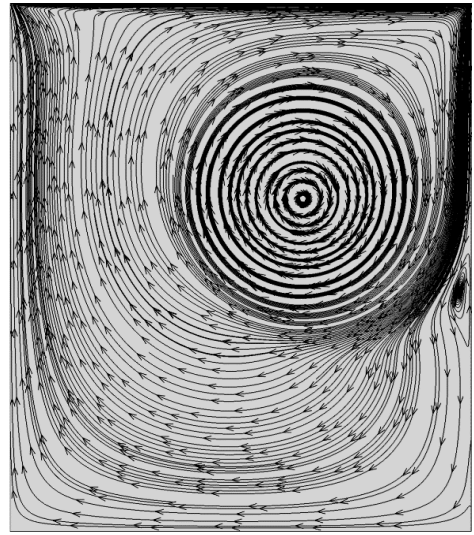
(b) 60000 Step

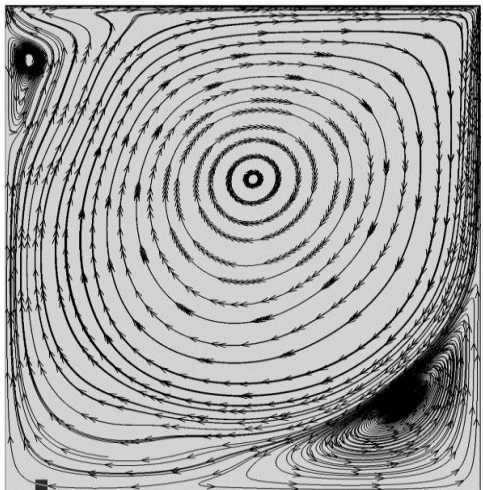
(c) 120000 Step

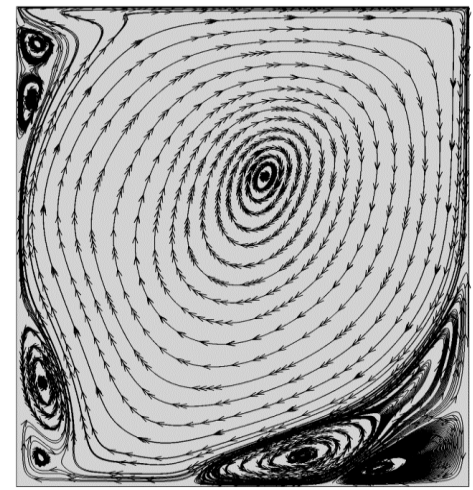
(d) 240000 Step

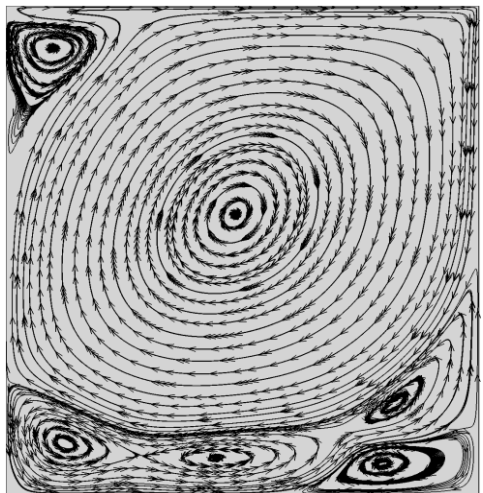
(e) 360000 Step

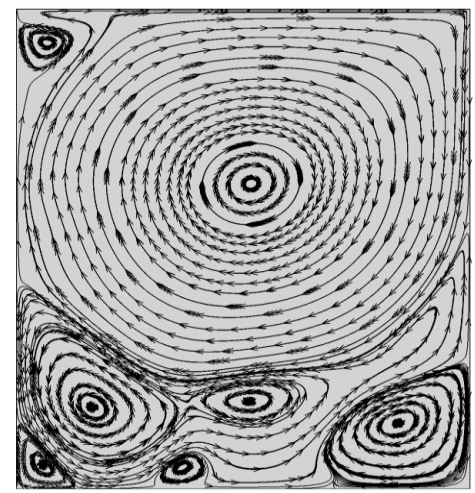
(f) 2000000 Step

Fig.3 Streamlines for the rectangular cavity flow with Ar=1.2, Re=6000 and different evolution step
(a) 20000S; (b) 60000S; (c) 120000S; (d) 240000S; (e) 360000S; (f) 2000000S

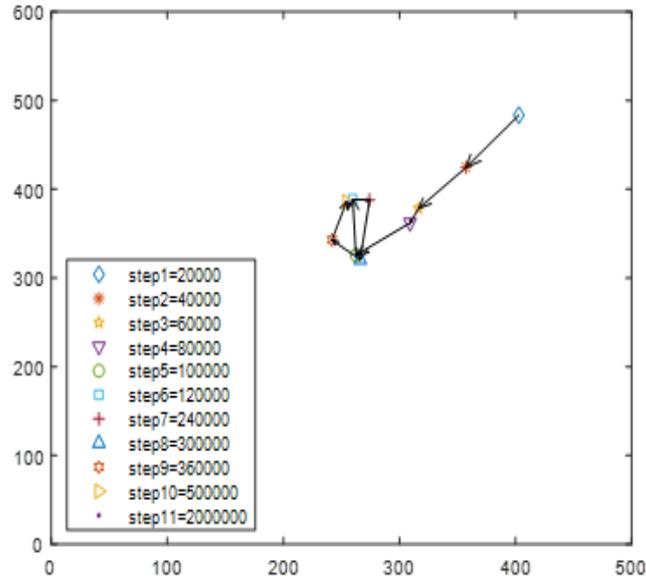

Fig.4 The center location of primary vortex varied with different evolution steps, Re=6000, Ar=1.2

Fig.4 is the location of center in the primary vortex varied with different evolution time (step). It isn't hard to see that firstly the center of the primary vortex begins to move down-wards by a big margin with the increase of evolution time, then the center of the primary vortex fluctuates near the center of the rectangular cavity at small amplitude with the extension of evolution time.

### (ii) The rectangular cavity flow with fixed Re while different Ar

Fig.5 corresponds to the contours of stream-function for cavities with fixed Re=6000 and different aspect ratio $0.5 \leqslant Ar \leqslant 2$, i.e., Ar =0.5, 0.8, 1.1, 1.2, 1.4, 1.6, 1.8 and 2.0, respectively. Fig.5 reveals that when Ar is smaller than 1, more likely, the rectangular cavity flow has a bigger secondary vortex along the horizontal direction. When Ar is bigger than 1, the lid-driven cavity flow consists of a series of vortexes at the bottom of the cavity, which superimpose on each other due to flow interactions. When Ar further improves until Ar is greater than the certain critical value, a secondary vortex with the similar size as the first big vortex just below the top lid appears at the bottom of the cavity. That is to say, at this time, there exist two bigger vortexes along the vertical direction of the cavity.

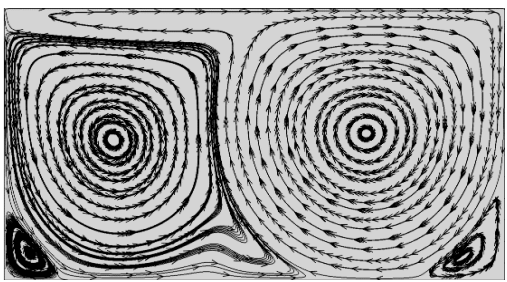
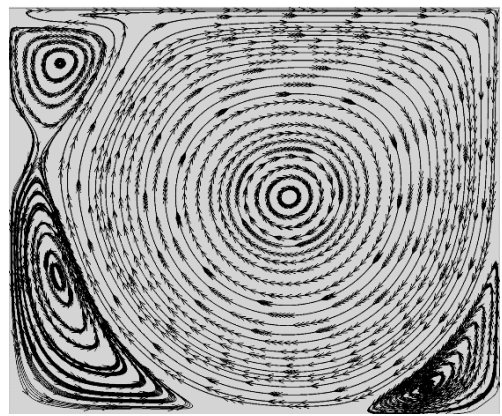

(a) Ar =0.5  (b) Ar =0.8

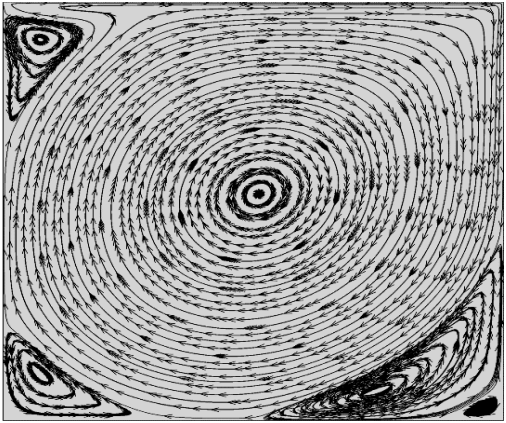

(c) Ar =1.1

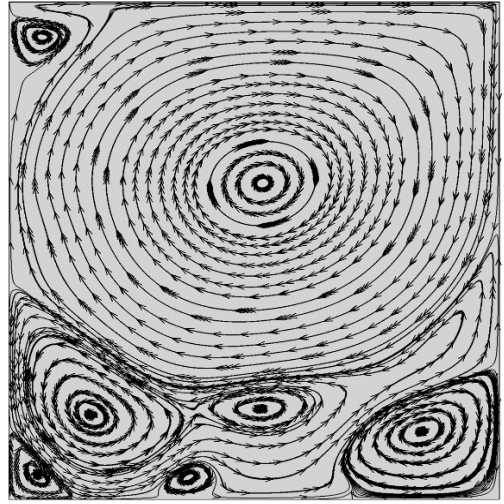

(d) Ar =1.2

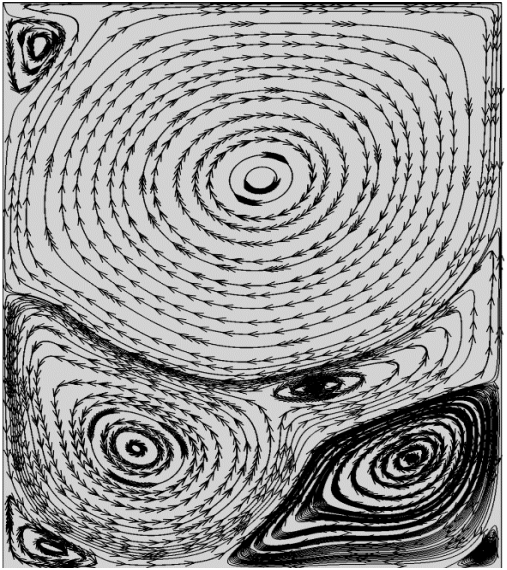

(e) Ar = 1.4

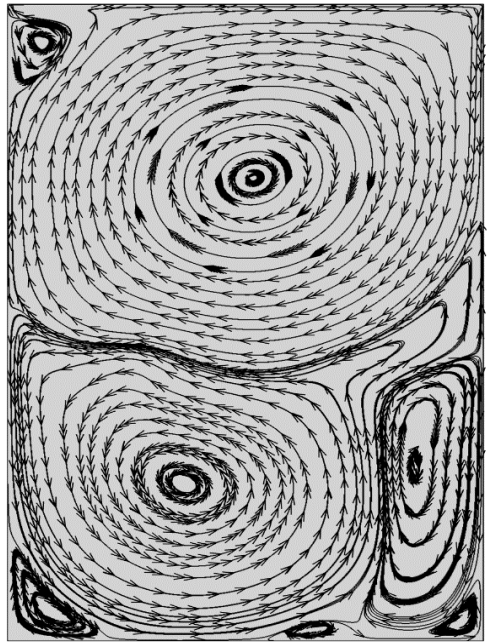

(f) Ar = 1.6

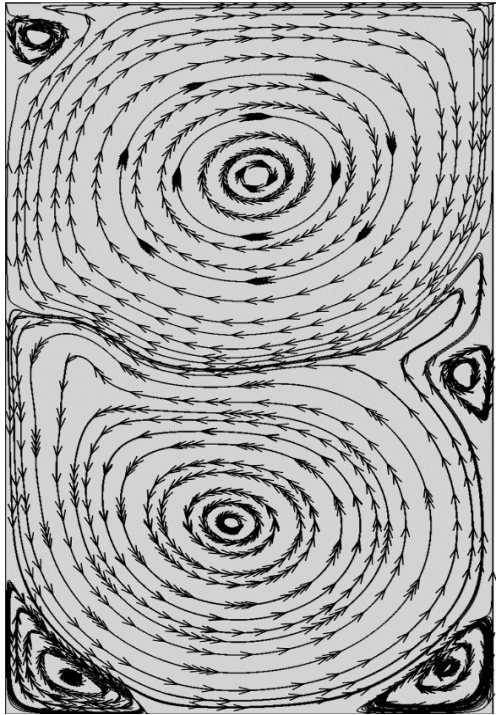
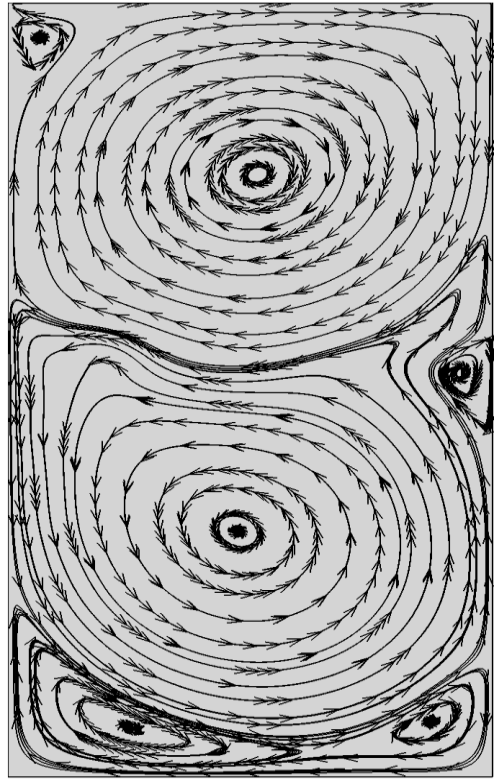

(g) Ar =1.8            (h) Ar=2.0

Fig.5: Streamlines for the rectangular cavity flow with fixed Re=6000 and different Ar.
(a) Ar =0.5; (b) Ar =0.8; (c) Ar =1.1; (d) Ar =1.2; (e) Ar =1.4; (f) Ar =1.6; (g) Ar =1.8; (h) Ar =2.0

### (iii) The rectangular cavity flow with fixed Ar while different Re

Fig.6 is the contour of stream-function for cavities with fixed Ar=1.2 under different Re, i.e. Re=4000, 4500, 5000, 6000, 6500 and 8000, respectively. Fig.6 indicates that several vortexes constantly generate, merge and disappear at the bottom of the rectangular cavity with the further increases of Re, namely, the contour of stream-function becomes more complex, the number and structure of eddies at the bottom of the rectangular cavity are varied with the change of Re.

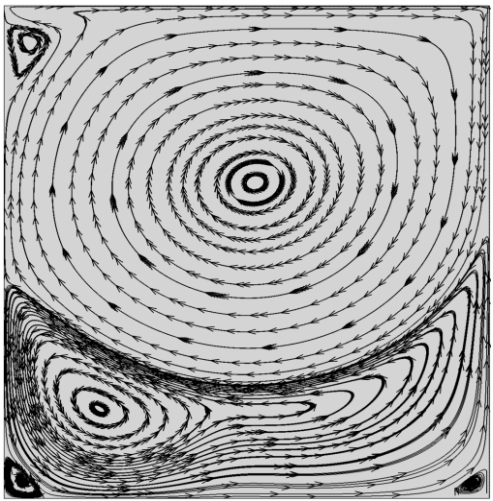
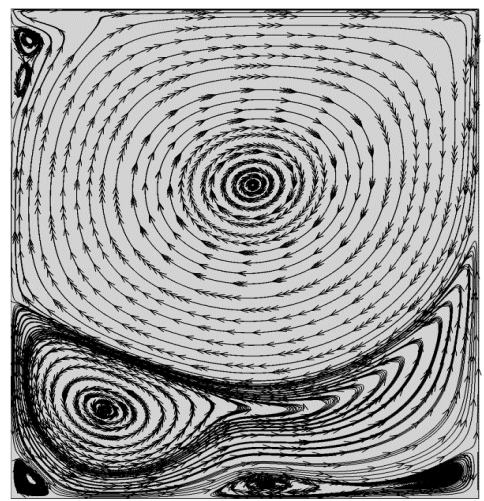

(a) Re=4000            (b) Re=4500

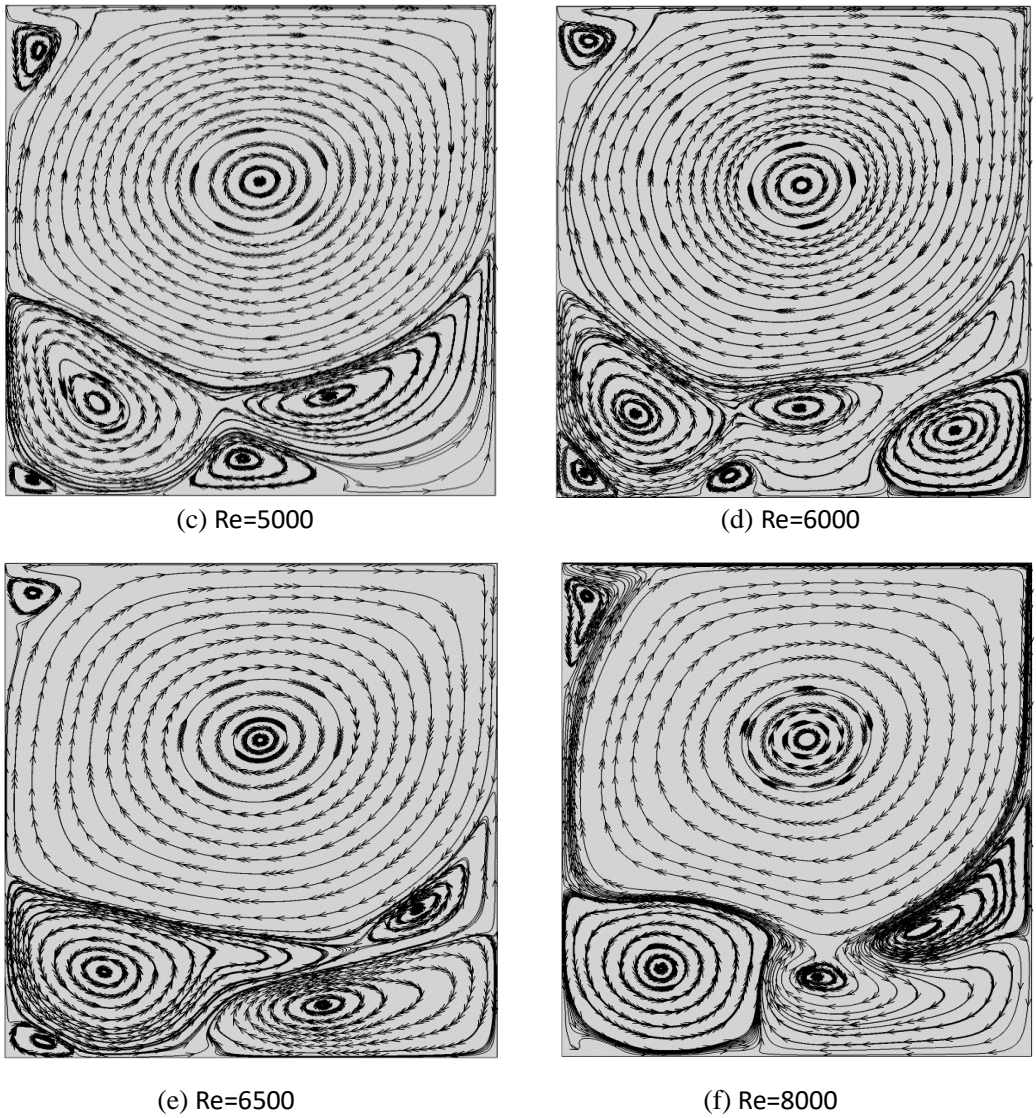

(c) Re=5000  (d) Re=6000

(e) Re=6500  (f) Re=8000

Fig.6: Streamlines for the rectangular cavity flow with fixed Ar=1.2 under different Re.

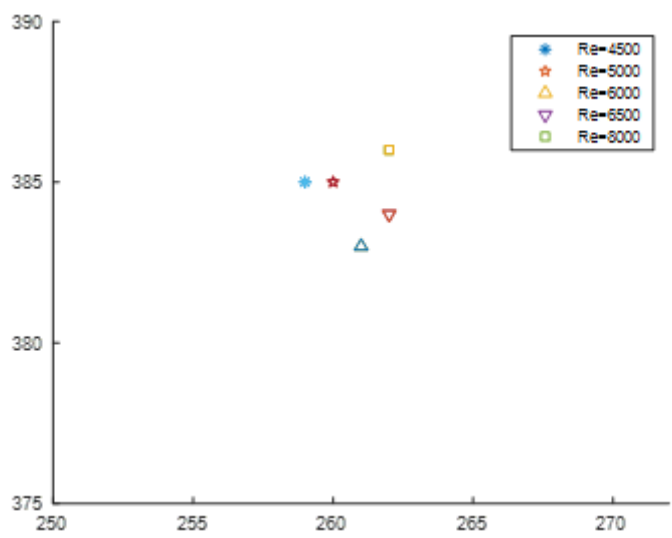

Fig.7  The center location of primary vortex varied with different Reynolds numbers, Ar=1.2

Fig.7 is the comparison of the center location of the primary vortex at different Reynolds numbers. Fig.7 demonstrates that the center of the primary vortex fluctuates near the center of the rectangular cavity at small amplitude with the change of Reynold number. Quantitative results listed in Table 2 also verify this conclusion. Additionally, the number of eddies at the bottom of the rectangular cavity shown in Fig.6 is already counted in the right column of Table 2.

Table 2 Stream function values, center positions of the primary vortex and the number of bottom vortex for different Re (Ar=1.2）

| Re | $\Psi_{max}$ | x | y | the number of bottom vortex |
|---|---|---|---|---|
| Re=4000 | fmax =5.9998 | 0.5180 | 0.7700 | 2 |
| Re=4500 | fmax =5.9382 | 0.5160 | 0.7680 | 3 |
| Re=5000 | fmax =5.8238 | 0.5180 | 0.7680 | 4 |
| Re=6000 | fmax =5.5420 | 0.5180 | 0.7640 | 5 |
| Re=6500 | fmax =5.3155 | 0.5220 | 0.7660 | 4 |
| Re=8000 | fmax =5.1616 | 0.5220 | 0.7700 | 3 |

## 3.3 Flow state of rectangular cavity varied with different Re and Ar

In this subsection, we utilize the observing point to analyze the flow state varied with different Re and Ar under the grid resolution of 500×(500*Ar). If Ar=0.4, U($x$=0.5, y=0.2) is selected as the observing point. If Ar=0.5, 0.6, U($x$=0.5, y=0.3) is regarded as the observing point, otherwise, U($x$=0.5, y=0.5) is taken as the observing point. $U_x$, $U_y$, respectively, denotes the flow speed in X-axis direction, Y-axis direction of the observing point.

## (i) Flow state of rectangular cavity with fixed Re while different Ar

Figs.8-10 are the $U_x$ value varied with different Ar under the same Re=6000, Re=6500, Re=8000, respectively. The X-axis in Figs.8-10 represents a certain period of the evolution time, which is between 1.5 million steps and 2.0 million steps. Y-axis indicates $U_x$ value. Note that only experimental results of even-number steps are cut out from the LBM simulation to save storage space in this paper. So evolution time between 1.5 million steps and 2.0 million steps is actually the time between 3.0 million steps and 4.0 million steps.

From Fig.8 we may get that when Re=6000 is fixed, the case where Ar=1.2 corresponds to the periodic state, the cases where Ar=0.5, 0.6, 0.8, 1.1, 1.5 correspond to the stable state, the case where Ar=1.6 corresponds to the aperiodic and unstable state.

From Fig.9 we may obtain that when Re=6500 is fixed, three cases where Ar=0.5, 0.6, 0.8 correspond to the periodic state, the case where Ar=1.1 corresponds to the stable state, while other three cases where Ar=1.2, 1.5, 1.6 correspond to the aperiodic and unstable state.

Fig.10 illustrates that when Re=8000 is fixed, one case where Ar=0.6 corresponds to the periodic state, one case where Ar=1.1 corresponds to the stable state, other cases where Ar=0.5, 0.8, 1.2, 1.5, 1.6 correspond to the aperiodic and unstable state.

Figs.8-10 interpret that the flow state of rectangular cavity with fixed Re constantly varies among the stable state, periodic state, aperiodic and unstable state, state transition is non-monotonical with the increase of Ar.

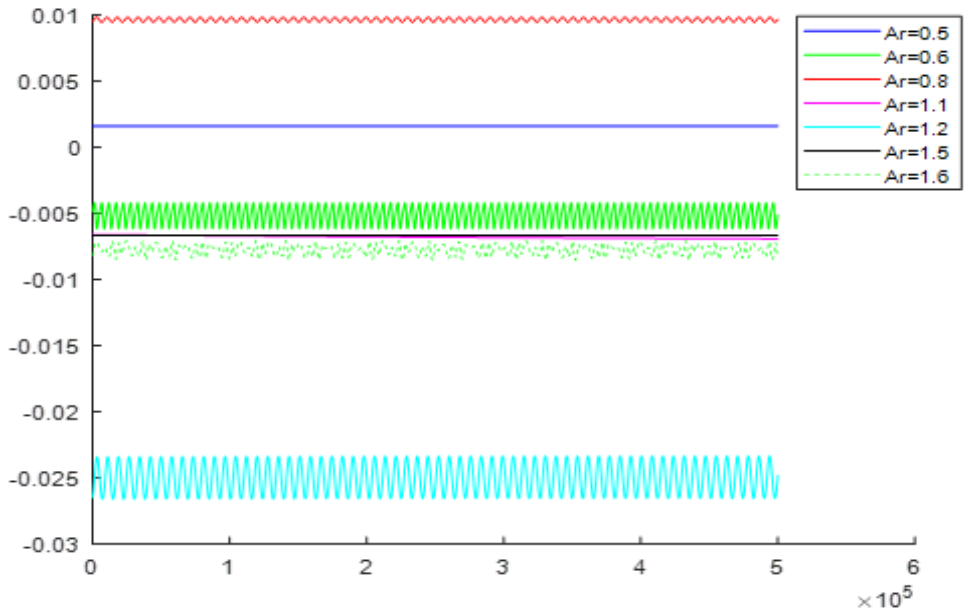

Fig.8 $U_x$ value varied with different Ar (Ar=0.5, 0.6, 0.8, 1.1, 1.2, 1.5, 1.6) under the same Re=6000

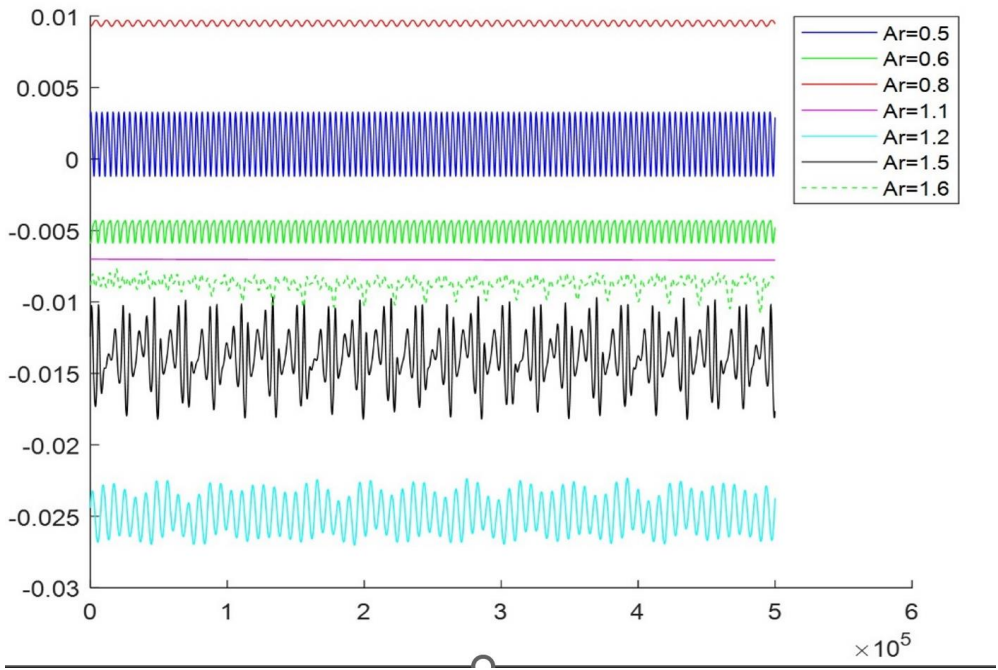

Fig.9 $U_x$ value varied with different Ar (Ar=0.5, 0.6, 0.8, 1.1, 1.2, 1.5, 1.6) under the same Re=6500

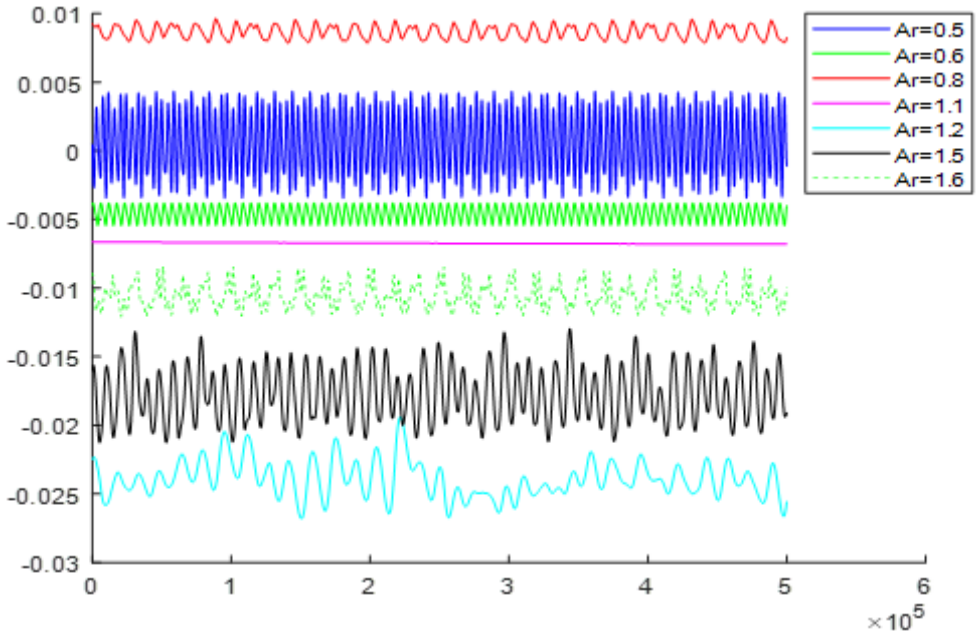

Fig.10 $U_x$ value varied with different Ar (Ar=0.5, 0.6, 0.8, 1.1, 1.2, 1.5, 1.6) under the same Re=8000

### (ii) Flow state of rectangular cavity with fixed Ar while different Re

Figs.11, 13, 15 are the $U_x$ value varied with different Re under the same Ar=0.5, Ar=0.8, Ar=1.2, respectively. The X-axis in Figs. 11, 13, 15 represents the evolutionary time from step 0 to step 2.0 million. Y-axis in Figs. 11, 13, 15 denotes $U_x$ value. Figs.12, 14, 16 are the phase diagram of the observing point varied with different Re under the same Ar=0.5, Ar=0.8, Ar=1.2, respectively. The X-axis, Y-axis in Figs. 12, 14, 16, respectively, indicate the $U_x$ value, $U_y$ value of the observing point selected by above method from step 1.9 million to step 2.0 million.

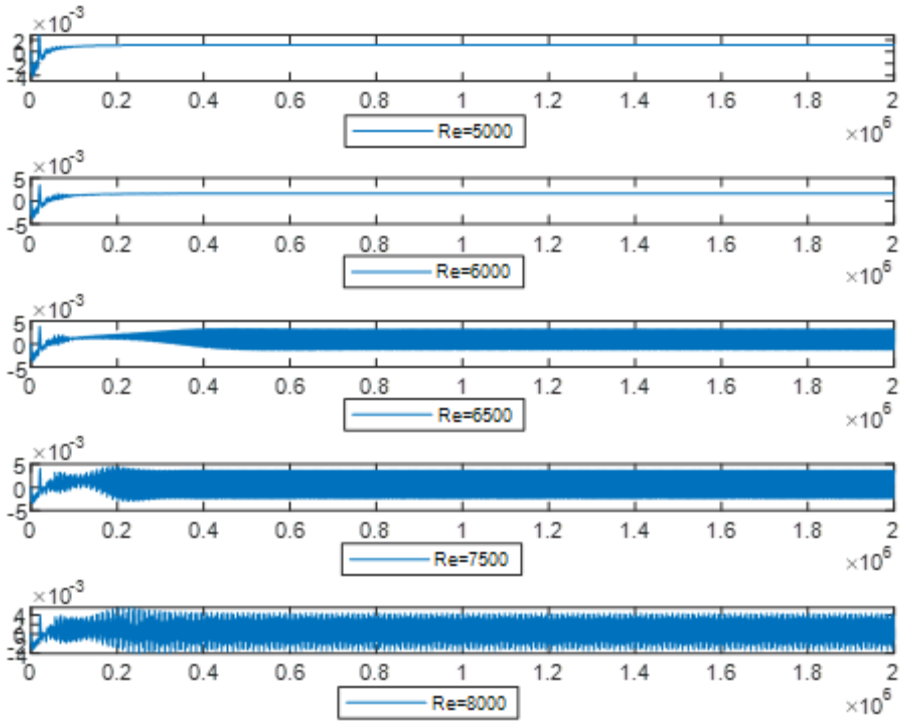

Fig.11   $U_x$ value varied with different Re under the same Ar=0.5

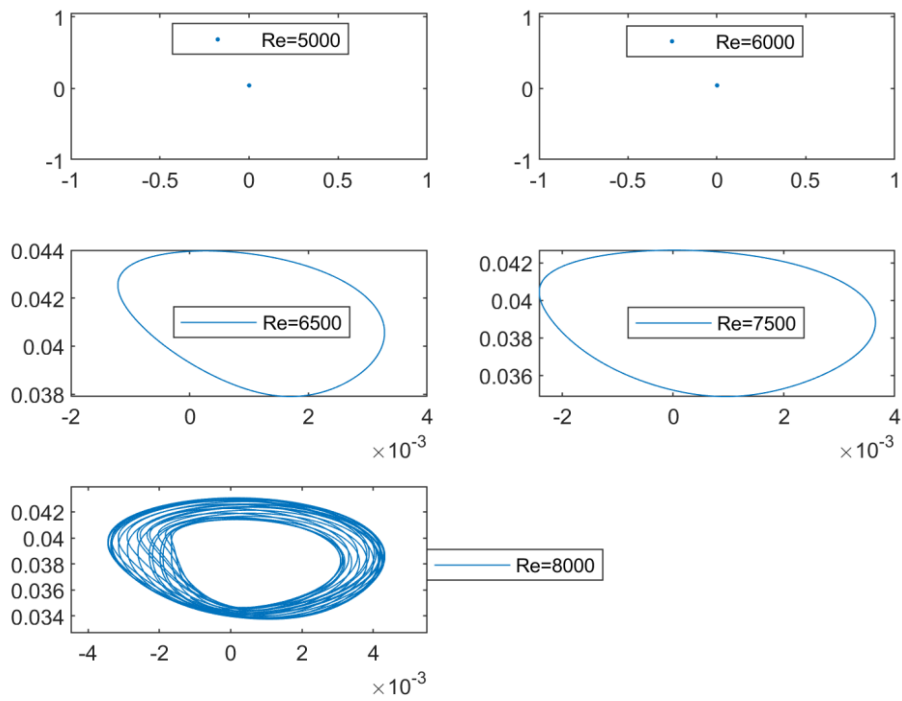

Fig.12  $U_x$-$U_y$ phase diagram varied with different Re under the same Ar=0.5

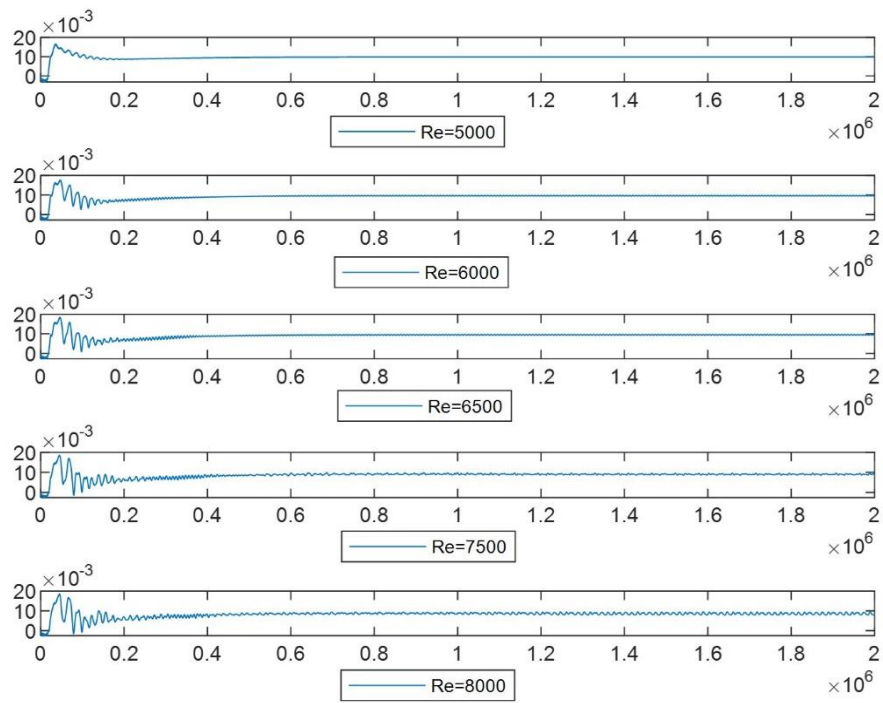

Fig.13  $U_x$ value varied with different Re under the same Ar=0.8

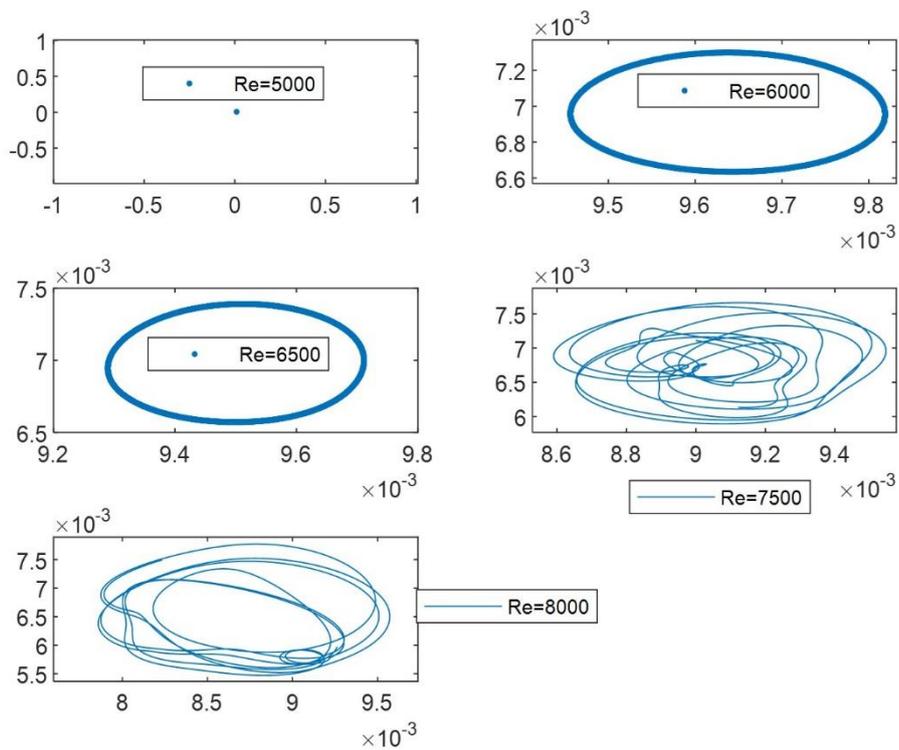

Fig.14 $U_x$-$U_y$ phase diagram varied with different Re under the same Ar=0.8

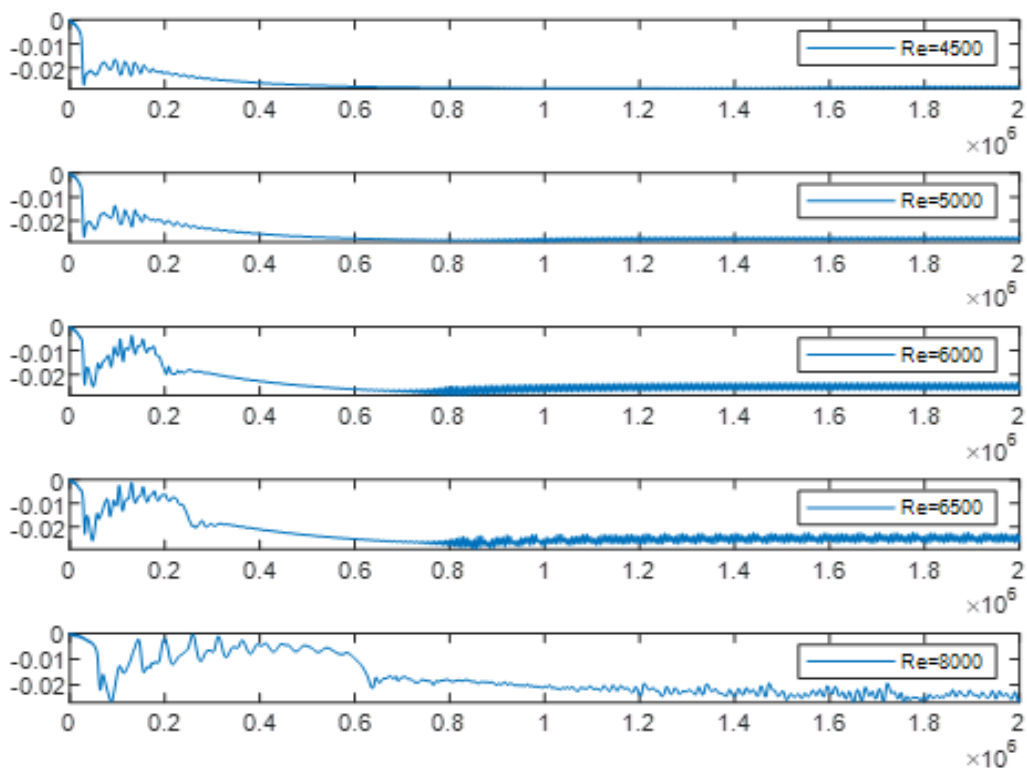

Fig.15 $U_x$ value varied with different Re under the same Ar=1.2

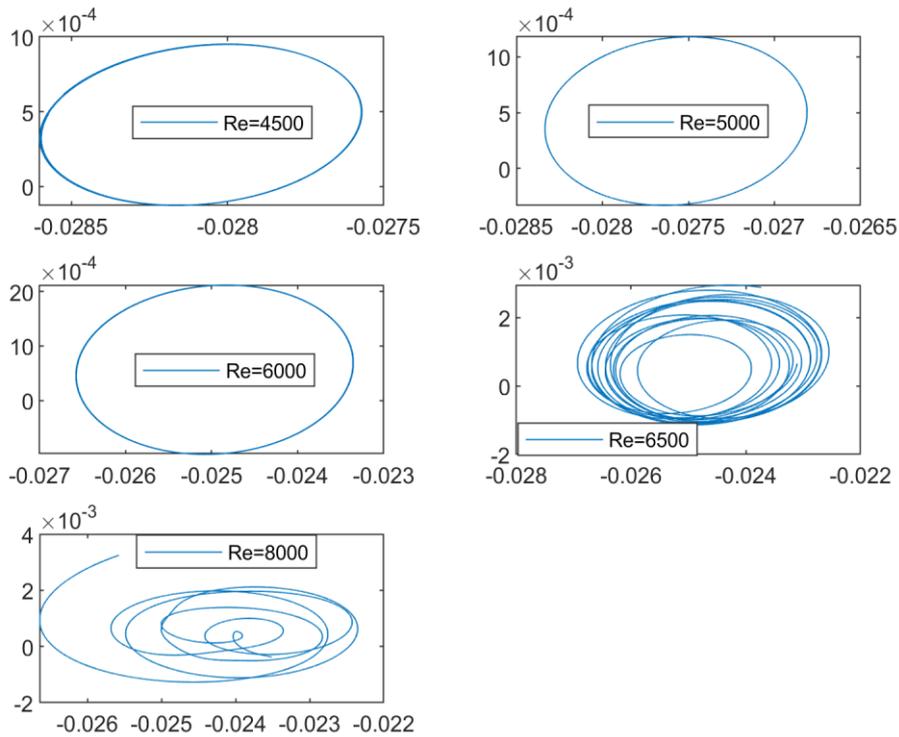

Fig.16　$U_x$-$U_y$ phase diagram varied with different Re under the same Ar=1.2

Figs.11, 12 demonstrate that the rectangular cavity flow changes from stable state, periodic state to aperiodic and unstable state with the increase of Re which varies from Re=5000, 6000, 6500, 7500 to Re=8000 while Ar=0.5 is fixed. Figs.13, 14, Figs.15, 16 also arrive at similar results while Ar=0.8, Ar=1.2 is fixed, respectively. In other words, the number of stable state becomes less, and the rectangular cavity flow gets more unstable when the increase of Re, this conclusion may also be drawn from the comparison of Figs.8-10.

### (iii) State statistics of the rectangular cavity flow with different Re and Ar

Finally, the flow states in the rectangular cavity with different Re and Ar are summarized and drawn in Fig.17, which validates that the evolutionary states are significantly dependent on Ar and Re. From the horizontal point of view, the flow state in the rectangular cavity with a fixed Ar (take Ar =0.8 or Ar =1.2 for example) changes monotonically from stable state, periodic state to aperiodic and unstable state when Re varies from 4000, 4500, 5000, 5500, 6000, 6500, 7000, 7500 to 8000. From the vertical perspective, the flow state in the rectangular cavity with a fixed Re varies non-monotonically with the increase of Ar. For instance, when Re=6000 is fixed while Ar changes from 0.4, 0.5, 0.6, 0.7, 0.8 to 1.0, the flow state varies from stable state, periodic state to stable state. When Re=6000 is fixed while Ar changes from 1.0, 1.2, 1.6, 1.8 to 2.0, the flow state constantly varies among the stable state, periodic state, aperiodic and unstable state. From lower left corner to upper right corner in Fig.17, the flow states change from the stable state, periodic state to aperiodic and unstable state with the increase of Ar and Re. Additionally, the stable state is mainly concentrated nearby Ar =1.0 in Fig.17, while the periodic state is mainly concentrated nearby Ar =0.6 and Re=6500 or Ar =1.3 and Re=5500 in Fig.17.

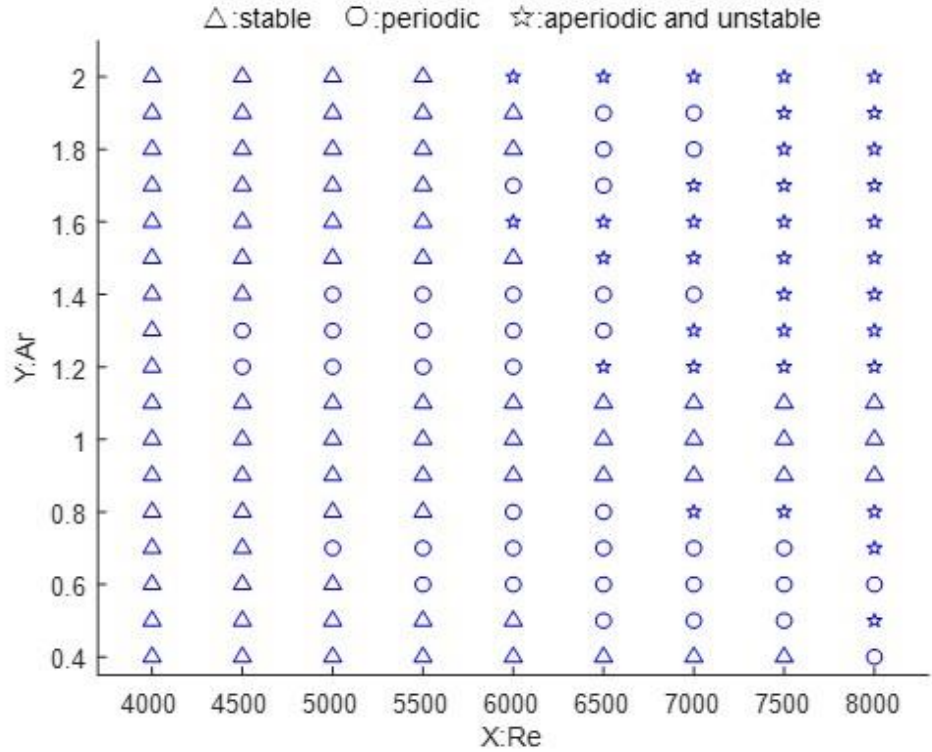

Fig.17    Flow state varied with different Re and Ar

Worthy of note that for the periodic state denoted by symbol ○ in Fig.17, the periods at different Ar and Re correspond to oscillation patterns with various amplitude and cycle length, as displayed in Figs.8-10. In order to the quantitative comparison of different periodic states, partial results about the length of period are listed in Table 3, the time data in Table 3 is measured in seconds. Table 3 shows that the length of period state in the rectangular cavity with a fixed Re varies non-monotonically with the increase of Ar, this is consistent with Fig.17. Moreover, the length of period nearby the stable region Ar =1.0 is comparatively longer, the length of period nearby both ends Ar =0.5, Ar =1.8 is comparatively shorter.

Table 3 The length of period varies with the change of Re and Ar

| Time(s) =step*dt  Re / Ar | 4000 | 4500 | 5000 | 5500 | 6000 | 6500 | 7000 | 7500 | 8000 |
|---|---|---|---|---|---|---|---|---|---|
| 0.4 | / | / | / | / | / | / | / | / | 13.14 |
| 0.5 | / | / | / | / | / | 16.26 | 16.58 | 16.86 | / |
| 0.6 | / | / | / | 20.20 | 21.08 | 21.67 | 22.13 | 22.46 | 22.73 |
| 0.7 | / | / | 24.26 | 25.06 | 25.92 | 26.82 | 27.58 | 28.18 | / |
| 0.8 | / | / | / | / | 30.23 | 30.85 | / | / | / |
| 1.2 | / | 30.31 | 30.30 | 30.63 | 31.16 | / | / | / | / |
| 1.3 | / | 30.02 | 29.89 | 30.37 | 31.46 | 23.92 | / | / | / |
| 1.4 | / | / | 29.90 | 29.51 | 23.23 | 23.69 | 24.06 | / | / |
| 1.7 | / | / | / | / | / | 17.12 | 17.58 | / | / |
| 1.8 | / | / | / | / | / | / | 17.73 | 18.12 | / |
| 1.9 | / | / | / | / | / | / | 18.15 | 18.68 | / |

## 4. Conclusions

As a very well-known benchmark problem, the lid-driven cavity flow is often simulated by numerical methods, such as finite difference, finite element, finite volume, Lattice Boltzmann Method (LBM), etc. In this paper, LBM is applied for the simulation of lid-driven flow not only in a square cavity but also in a rectangular cavity with aspect ratios of 0.4–2.0 and Reynolds numbers of 4000–8000, almost all the physical details of this flow and the dynamic features of the vortices are well captured by these simulations. The vortex structure of the flow, such as the stream function values and center positions of the primary and second vortexes are investigated and compared with previous findings, and the flow state of the rectangular cavity for different Ar and Re are analyzed and summarized in detail. Some interesting phenomenon is discovered and main results are obtained.

(i) The number and structure of the vortexes varies with the enlargement of Ar and Re. The size and center position of the primary eddy just below the top lid are affected by the Reynolds number and evolution step, but is not so much by the cavity depth.

(ii) The vortex structure with fixed Ar appears more complex as Re increases. The eddies at the bottom of the rectangular cavity evolve continuously to form a series of vortexes as the cavity Ar is increased above a critical value. When Ar is greater than 1, the lid-driven rectangular cavity flow contains several vortexes which superimpose on each other due to flow interactions.

(iii) The evolutionary state of the rectangular cavity flow is also closely related with Ar and Re. The flow in the rectangular cavity with a fixed Ar changes monotonically from stable state, periodic state to aperiodic and unstable state with increasing Re, while the flow state in the rectangular cavity with a fixed Re varies non-monotonically with the increases of Ar. In addition, the cycle length of periodic states varies with the change of Ar and Re.

The interesting results found in this paper may provide reference for some engineering design and practical application. Future efforts need to extend these simulations to the fully three-dimensional flow with different boundary conditions.

## Acknowledgment

This investigation is supported by the State Scholarship Fund of China Scholarship Council under Grant No 201806415025, and supported by the National Natural Science Foundation of China under Grant No 61202051